\documentclass[review]{elsarticle}
\usepackage{lineno,hyperref}
\usepackage[utf8]{inputenc}
\usepackage[english]{babel}
\usepackage{natbib}
\usepackage{xcolor}

\newcommand{\hlnewtwo}[1]{\textcolor{black}{#1}}
\newcommand{\hlnew}[1]{\textcolor{black}{#1}}
\newcommand{\hlcolor}[1]{\color{black}}

\usepackage{amssymb}

\journal{International Journal of Human - Computer Studies}

\begin{document}

\begin{frontmatter}



\title{Imagining Future Digital Assistants at Work:\\A Study of Task Management Needs}


\author[label7]{Yonchanok Khaokaew \fnref{lbequ}}
\author[label1]{Indigo Holcombe-James\fnref{lbequ}}
\author[label2]{Mohammad Saiedur Rahaman\fnref{lbequ}}
\author[label8]{Jonathan Liono\fnref{jonath}}
\author[label2]{Johanne R. Trippas}
\author[label2]{Damiano Spina}
\author[label4]{Peter Bailey\fnref{peter}}
\author[label3]{Nicholas Belkin}
\author[label5]{Paul N. Bennett}
\author[label2]{Yongli Ren}
\author[label2]{Mark Sanderson}
\author[label2]{Falk Scholer}
\author[label5]{Ryen W. White}
\author[label7,label2]{Flora D. Salim \corref{cor1}}

\cortext[cor1]{Corresponding author\\   Email addresses:  flora.salim@unsw.edu.au (Flora D. Salim)}


\fntext[lbequ]{Equally contributing first authors}
\fntext[jonath]{This work was done when Jonathan Liono was employed at RMIT University.}
\fntext[peter]{This work was done when Peter Bailey was employed at Microsoft.}
\address[label7]{School of Computer Science and Engineering, University of New South Wales (UNSW), Sydney, NSW, Australia.\fnref{label7}}
\address[label1]{School of Media and Communication, RMIT University, Melbourne, VIC, Australia.\fnref{label1}}
\address[label2]{School of Computing Technologies, RMIT University, Melbourne, VIC, Australia.\fnref{label2}}

\address[label8]{PT. Global Tiket Network (tiket.com), Singapore \fnref{label7}}

\address[label4]{Canva, Sydney, NSW, Australia.\fnref{label4}}
\address[label3]{School of Communication and Information, Rutgers University, New Jersey, USA.\fnref{label3}}


\address[label5]{Microsoft Research, Redmond, USA.\fnref{label5}}

\begin{abstract}
Digital Assistants (DAs) can support workers in the workplace and beyond. However, target user needs are not fully understood, and the functions that workers would ideally want a DA to support require further study. A richer understanding of worker needs could help inform the design of future DAs. We investigate user needs of future workplace DAs using data from a user study of 40 workers over a four-week period. Our qualitative analysis confirms existing research and generates new insight on the role of DAs in managing people's time, tasks, and information. Placing these insights in relation to quantitative analysis of self-reported task data, we highlight how different occupation roles require DAs to take varied approaches to these domains and the effect of task characteristics on the imagined features. Our findings have implications for the design of future DAs in work settings and we offer some recommendations for reduction to practice.

\end{abstract}




\end{frontmatter}

\biboptions{authoryear}
\bibliographystyle{elsarticle-num-names} 
\section{Introduction}

Early digital assistants (DAs) were developed to support people by assuming and completing their tasks \citep{bentley2018understanding}. 
In this article, we focus specifically on the use of DAs as a tool to support the productivity of participants \citep{mcgregor2017more,kocielnik2018designing} in work settings. 
DAs are typically oriented towards supporting users by helping them complete tasks \citep{thies2017you} and retrieve requested information \citep{grudin2019chatbots}. In a work context, existing research has examined the potential for DAs to schedule meetings \citep{cranshaw2017calendar}, manage to-do lists \citep{gil2008towards}, recommend applications \citep{ProCosem}, streamline email inboxes \citep{faulring2010agent}, seek information \citep{liao2016can}, and take notes during meetings \citep{mcgregor2017more}. However, while significant research effort has focused on the use of \citep{mehrotra2017hey}, and user satisfaction with, existing DAs \citep{gebauer2008user}, comparatively little attention has been paid to what target users might desire beyond what is already available. As a result, the functions that workers would actually imagine a future DA could support remain underexplored.

This article reports on a study in the context of task management in work settings. We use qualitative data collected during a user study of 40 worker participants that aimed to understand participant work tasks in their regular work-life quantitatively. We intended to use the quantitative data generated from the user study to develop a DA that could intelligently support these same tasks. \hlnewtwo{Such as a DA would need to be able to fulfill a wide range of tasks on a regular basis, both due to the complexity of work in general, as well as because our study had a wide range of work tasks. Therefore, to keep the broad scope of DA, the definition of DA in this work is a software agent that can perform tasks or services for supporting the user regardless of the input and platform.}
This quantitative data was gathered via a multi-staged process of contextualised data collection. Over four weeks, we recorded self-report data on the work tasks undertaken by each participant resulting in a record of 4,309 tasks. At regular intervals, participants were prompted to manually annotate these tasks through an Experience Sampling Method (ESM) and a Daily Reconstruction Method (DRM) \citep{liono2019building}. \hlnew{In addition, to capture the context-rich activities and behaviours of our participants in a comprehensive manner, we devised a logging procedure that facilitates the smartphone-based collection of the contextual sensor signals associated with different tasks.}  
At the end of each week, individual participants met with a member of the research team and were asked to reflect on the preceding week's tasks and to describe how an imagined DA could have assisted. By doing so, these imaginings were contextualised by the quantitative task data we had collected. Each participant imagined the features of a future DA in direct response to their most recent work experiences. We ask: what can our data tell us about the tasks and activities workers imagine a future workplace DA supporting? \footnote{Note that we use the term ``workplace'' to mean work settings in general, not only a physical work environment.}

This article sets out to answer the following three research questions (RQs): 
\begin{itemize}
    \item \textbf{RQ1}: When contextualised by their recently-performed tasks, what features do workers imagine a DA for the workplace should take on? 
    \item \textbf{RQ2}: Do these imagined features differ across occupation roles? 
    \item \textbf{RQ3}: Do these imagined features relate to the tasks performed by participants?
\end{itemize}

In doing so, we respond to the recent call from \citet{maedche2019ai} for research that investigates people's needs of DAs. In that call, the authors suggest that a goal of research in this area should be to develop nuanced understandings of what different users, operating in different contexts, expect of DAs. In investigating what features workers would imagine a future DA for the workplace taking on, we offer a contribution to these broader efforts.

We answer our research questions in two stages. The first stage focuses specifically on our qualitative interview data, gathered weekly in conjunction with the quantitative ESM and DRM data collected in the preceding week as the proxy for stimulated recall. We analyse this through thematic analysis that revealed workers imagined a workplace DA that manages user time, tasks, and information. We then look to understand if and how these imagined features differed by occupation role, revealing clear differences in occupation and also by aspects of the tasks being undertaken. We {identify three practical functions to guide future developments in workplace digital assistants}. By bringing together qualitative and quantitative data and analysis, we {take a step towards how we might better understand user needs of future DAs}. Our user study is uniquely situated in the personal task management sphere, with the aim of capturing what users imagine a DA could do to help them in their work tasks.

Our study confirms existing research that discusses DAs managing users' time, tasks, and information. It also extends existing literature by focusing on user desires rather than technical possibilities, and contextualising these desires within existing user work tasks. Further, by placing these qualitative insights in relation to quantitative analysis of both self-reported task data and the results of the qualitative thematic analysis, we identify connections between occupation roles and work tasks and the capacities they imagine a workplace DA could assume. Our article specifically contributes to the field by extending the existing literature in at least the following ways: 

\begin{itemize}
    \item Demonstrating the perceived utility of DAs that support user time, tasks, and information by target users.

    \item Highlighting how different occupation roles require varied approaches to each of these support domains. 
    
    \item Illustrating the effect of task characteristics on the DA features that users desire.
    
    \item Providing implications for the future design of DAs that utilise these imagined features.
    
    \end{itemize}

The remainder of the article is organised as follows. Section 2 provides an overview of existing research on DAs, including within work settings. Section 3 presents the methods used and the two stages of analysis that were undertaken. Section 4 outlines the empirical results in the broader context. Section 5 discusses implications for the design for DAs and the limitations of the study methodology. Finally, Section 6 concludes the article and outlines future work. 

\section{Related work}
Previous work in several areas is related to our research, including digital assistants in general and in the workplace, and envisioning digital assistants for work.
\subsection{Digital Assistants }
The development of DAs can be traced back to the 1980s and 1990s, when products such as Apple's Knowledge Navigator and AT\&T's PersonaLinks were developed to support user productivity \citep{lopatovska2018personification}. Contemporary, or next-generation \citep{meurisch2017reference}, DAs such as Cortana, Siri, and Alexa are often referred to as Conversational Agents (CAs) because interaction is generally via speech or text \citep{motalebi2019can,feng2020my,meyer2020chatbots,chatbottochi,voiceasselderly,TERSTAL2020102409,GILBERT201530, zamani2022conversational}. The CA then responds on the basis of predefined commands (such as ``send text message'') \citep[p.~2208]{mcgregor2017more}. The CAs topic has received a lot of attention from researchers and many research articles have been published on this subject \citep{DEBARCELOSSILVA2020113193,RAPP2021102630}. CAs are referred to by most of these articles as Intelligent Personal Assistant (IPA) or DAs. Our participants, however, described a DA that did not necessarily require input from the user (verbally or otherwise), but rather inferred instructions from contextual clues. Due to the lack of knowledge about how users perceive DAs, we aim to address this question in this study.

\subsection{Workplace Digital Assistants}
Having long played a role in supporting the completion of user work tasks, researchers have considered how DAs might also automate user scheduling practices and, in doing so, improve user productivity~\citep{cranshaw2017calendar}. Likewise, \citet{refanidis2010constraint} discuss the possibility of a DA that both automates scheduling and learns from user preferences to improve productivity. In addition, DAs might help workers avoid the information overload from various sources of information, as \citet{SOROYA2021102440} suggest that workers would feel information anxiety since it is strongly associated with information overload. Others, such as \citet{ludford2006because}, and \citet{kamar2011jogger}, have investigated how DAs might support users by providing context- and location-specific reminders. \citet{faulring2010agent} outline the possibilities for DAs to improve productivity by managing users' email inboxes. Similarly, \citet{freed2008radar} describe a DA that identifies tasks to be completed arising from received information (such as messages or email). In turn, \citet{myers2007intelligent} describe a DA that supports the user by providing task and time management,  \citet{LI2019704} mention the DA used in the library to support users and help them to complete their tasks, and \citet{mcgregor2017more} describe a DA that contributes to user productivity by taking notes during meetings. Including a recent study by \citet{BELKADI2020105732}, which proposes an intelligent assistant system for supporting workers by providing the right information at the right time and in the appropriate format regarding their context. Although their work explains a clear concept of components and features of a decision-making system, these features are only applicable to the systems used in the manufacturing industry.

However, the tasks that workers want a workplace DA to support remain underexplored. As \citet{meurisch2017reference} argue, this lack of insight into what users actually want from a DA impedes their development. There is considerable practical value in user studies that ask participants to reflect on, and imagine, a DA that would directly benefit them and how that DA would do so. Although research by \citet{luger2016like} and \citet{ToCHIuseVA} is useful in that they ask participants how they actually use their DA/CA with the intention of informing future iterations, we take a different approach. Rather than examining how workers use existing DAs, we investigated the tasks workers imagined a workplace DA might help with in the context of their existing work tasks.

\subsection{Workplace Digital Assistants Imagined by Workers}

\hlcolor

In the last decade, there have been several studies of user demands about intelligent assistants in the workplace. \citet{mcgregor2017more}, for example, studied a speech-based agent system for use in a group meeting setting. They investigated how such a intelligent system might perform in the collaborative work setting and what users might respond to it. \citet{adamczyk2005method} provide useful insight into the issue of negotiating digital distraction, they do so by evaluating participants' pupil size to determine their workload, and therefore the appropriateness of an interruption. Other works by \citep{meyer2020chatbots, feng2020my} attempted to investigate and identify application areas that might benefit employees in digital workplaces by using interviewing and a systematic literature review method. \citet{afzali2018cellphone} observed their participants' experience of digital distraction to understand how a DA might assist in resolving such distractions. Likewise, \citet{czerwinski2000instant} investigated how DAs might assist the user in negotiating digital distraction by asking participants to complete tasks with the interruption from the instant message program. Research by \citet{myers2007intelligent} studied how a DA that might assist users by managing their time and tasks. The authors asked participants to use their implemented program to perform various use cases and asked users for feedback on this implemented program usage via a Belief-Desire-Intention agent system.

In spite of this, there are still some gaps in these prior studies that we hope to fill in this study. Firstly, although these previous studies have employed different methods to better understand users' needs regarding their workplace DAs, there are no previous studies that have examined contextual data derived from their existing work tasks as well as having the users reflect on their needs informed by these contextual data. This contextual data containing the task information should help participants to determine which form of support they need from the DAs in their working environment. Additionally, previous research focuses on specific types of DAs; for instance, the work by \citep{meyer2020chatbots, feng2020my} focuses only on conversational agents for use in the workplace. Our aim is to understand user requirements without being restricted by the type of DA. Lastly, the occupation roles of users which have a strong relationship with their tasks, are not considered much in prior work. For example, research by \citet{mcgregor2017more} and \citet{myers2007intelligent} studied only the needs of participants in management roles. As we investigate the needs of participants with regard to their tasks, we will also examine whether those needs are different across positions. 

\color{black}

\section{Method}

\begin{figure}[!t]
  \centering
  \includegraphics[width=\linewidth]{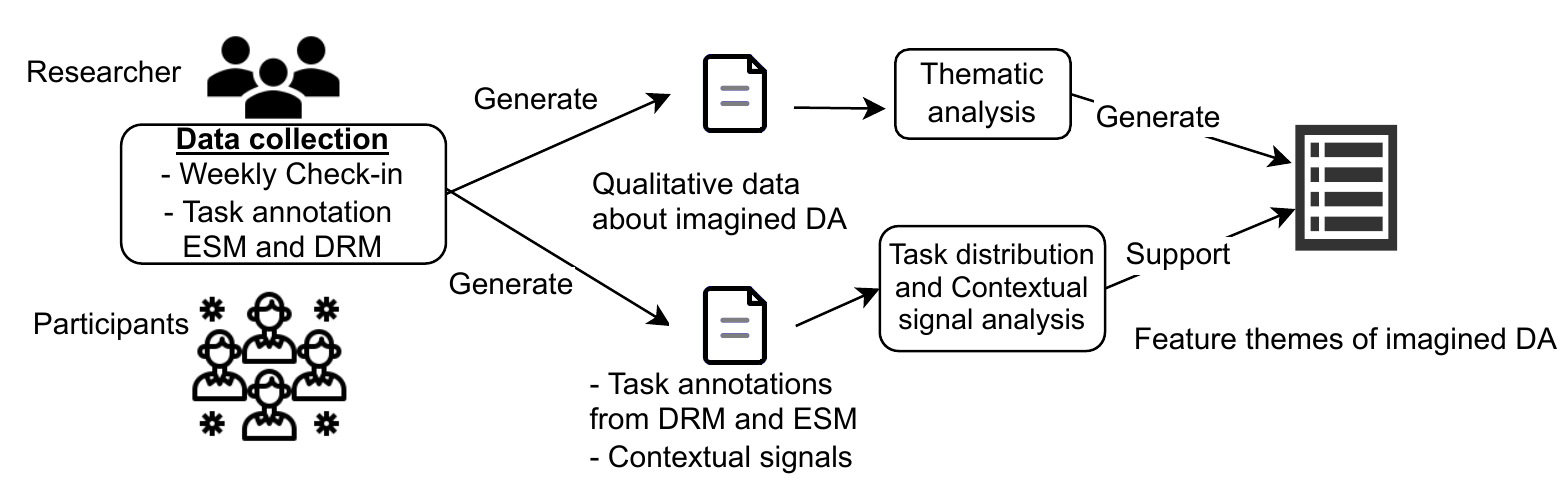}
  \caption{Overview of the approach for data collection and analysis used in the study.}
  \label{fig:overview-model}
\end{figure}

In this section, \hlnew{we provide an overview of the data collection and analysis methodology. It should be noted that this data collection is a part of a broader project which aims to understand how tasks progress over time, enabling DAs to help with current activities and support future activities.}
Firstly, we collect self-reported task data through the Experience Sampling Method (ESM) ~\citep{hektner2007experience} and the Daily Reconstruction Method (DRM)~\citep{czerwinski2004diary} to form a context-rich task data set that helps us characterise the daily tasks performed by users across different occupations. \hlnew{The ESM/DRM method was chosen over speculative design activities given their greater ecological validity \citep{verhagen2016use}. ESM, which provides a comprehensive view of an individual's daily life, is able to provide insight into how tasks develop over time. The DRM method is designed to capture a full picture of a day, including the duration of daily activities, without disrupting daily life \citep{bylsma2011emotional}. DRM can provide complementary information about a person's activities on a daily basis. By using these two methods, the participant will be able to understand what type of support they will need when they review their task annotations.} The contextual and auxiliary signals were continuously logged over a four-week period from 6:00 AM to 7:00 PM on weekdays. \hlnew{We did not wish to record sensor data 24/7 (including their locations, which are anonymized) since doing so may be too intrusive for study participants. The collection of data was only done from Monday through Friday during normal working hours. Typically, Australian working hours are between 9:00 AM and 5:00 PM. While collecting sensor data, we set the recording to begin at 6:00 AM and end at 7:00 PM. This is to better enable the capture of the various user tasks that may be encountered during the process. Participants might work earlier in the morning and perform more physical activity (such as chefs) limiting their ability to self-report. Some participants may finish late, particularly busy professionals who work overtime to complete their work projects.}

At the end of each week, individual participants met with a member of the research team to ensure the sensing applications used to collect the self-report data and contextual signals were running correctly, and to ensure the consistency and accuracy of task annotations by participants. For example, unexpected activities or gaps were identified by the researcher and clarified through conversation. As part of this conversation, the participant was shown a spreadsheet containing their self-reports, and was asked to reflect on the preceding week's tasks and to describe how an imagined DA could have assisted with each of them. Notes taken by the researcher during these conversations were added to each participant's file, generating a corpus of qualitative data. This qualitative data about imagined DA's is the core data set analysed in this article, using both qualitative and quantitative analysis, in conjunction with the task annotations gathered through ESM and DRM.

\hlnewtwo{In order to analyse the collected data, we use two different stages. In the first stage, we focused exclusively on qualitative interview data collected weekly. We used thematic analysis to reveal that workers imagined a DA that manages their time, tasks, and information at work. We then examined whether and how these imagined features varied between occupation roles as well as between aspects of tasks performed. The second step of our analysis focused on determining whether the quantitative sensing data collected from participants could provide evidence that these features were in demand by users. Using contextual signals, we can divide our participants into clusters. Based on the characteristics of each cluster, we determine the potential demand that participants in each group may have. Then, this desire is compared to the needs of participants derived from thematic analysis, which was the outcome of the previous stage. In the following sections, we provide a detailed description of our methods for collecting and analysing data.}

\subsection{Data Collection}

\subsubsection{Participant recruitment}
\label{subsubsec:participant recruitment}
\hlcolor

We recruited 53 participants for our user study\footnote{\hlnew{The data collection protocol was reviewed and approved by the Human Research Ethics Committee at RMIT University, ref ``SEHAPP 09-18 SALIM-LIONO''.}} using a system called ORSEE (Online Recruitment System for Economic Experiments)\footnote{\url{http://www.orsee.org/web/}}. We also used online classified advertisements and community websites (for example, meetups, social media, and an alumni website). Potential participants completed an Expression of Interest form and were screened against a set of requirements which included: 
\textit{(i)} willingness to actively participate for four consecutive weeks; 
\textit{(ii)} ability to come to the intake session and weekly meetings; 
\textit{(iii)} being employed part- or full-time; and 
\textit{(iv)} owning an Android smartphone with version $\geq$ 5.0. 
\hlnew{Furthermore, the potential participants were given the participant information sheet and consent form before submitting the expression of interest. This sheet and consent form explained the study in layman's terms, including which data was collected. As part of the introduction to the study, the researchers went through the information sheet with each participant and were asked to sign the consent form if they agreed to take part. Participants could withdraw from the study at any point in time. We used four techniques to de-identify collected data, \textit{(i)} location randomization, \textit{(ii)} one-way hashing, \textit{(iii)} k-anonymity,  and \textit{(iv)} manual named-entity de-identification. All data is stored on secure servers.}
Participants were compensated with AUD \$600 for their contributions. The payment was divided into four installments, and the participants were paid after each weekly meeting, with the final payment being the largest to help ensure that they remained engaged for the full study duration. Note that all participants had Australia as country of residence and were fluent in English.

\begin{table}[t!]
\small
\centering
\hlcolor
\caption{Study participants and their occupation groups.}
\label{tab:professionals}
\resizebox{\textwidth}{!}{\begin{tabular}{| p{9cm} |p{4cm} |}
\hline
\textbf {Occupation group | Job Role Details} & \textbf{Participant ID}\\
\hline

\textbf{ Manager } | Project manager, Student service, Event manager, Strategy manager, Project manager, Business development manager, Start-up founder, Sales manager, Product owner, Account manager, Business development manager, Business owner & P004, P005, P021, P023, P026, P028, P033, P034, P039, P044, P048, P052 \\ \hline
\textbf{ Technicians and associate professionals } | Technical writer and editor, Graphic designer, Lab technician, Job recruiter & P016, P030, P032, P047 \\ \hline
\textbf{ Professionals } | Office worker, Principal advisor, Business analyst, Artist planning exhibition, Architect, Book editor, Nurse, Financial specialist, Marketing consultant, Structural engineer, Paralegal & P017, P019, P027, P031, P037, P038, P042, P043, P045, P049, P051 \\ \hline
\textbf{ ICT professional } | IT security analyst, Development relation specialist, Digital service analyst, Ux designer, Test analyst, Front end developer, Data specialist, Software developer, Software delivery lead & P018, P024, P025, P029, P035, P036, P040, P050, P053 \\ \hline
\textbf{ Clerks } | IT project coordinator, Library officer & P020, P022 \\ \hline
\textbf{ Service and Sales } | Chef, Sales and marketing & P041, P046 \\ \hline

\hline

\end{tabular}}
\end{table}

\color{black}

Since the aim of this article is to explore the tasks workers imagined a workplace task assistant should support, we consider only the 40 participants (25 males, 15 females) out of our initial pool of 53, who at the time of data collection identified as `workers'. In this context, workers were self-defined in relation to their work tasks. The different occupation groups our participants reported are listed in Table \ref{tab:professionals}. We categorise the occupation group of these participants by using the major groups from the International Standard Classification of Occupations (ISCO-88) \citep{hoffmann1995revised}. These major groups found in our dataset consist of \textit{Managers}, \textit{Professionals}, \textit{Service and sales workers}, \textit{Clerks}, and \textit{Technicians and associate professionals}. \hlnew{As mentioned by \citep{ICTdiff}, Information Technology (IT) jobs often require IT professionals to utilise unique skills, competences, and knowledge to successfully complete their jobs, and IT workers tended to perform significantly fewer citizenship behaviors (such as contributing to preventing a problem or participating constructively in political processes at work ) than colleagues in non-IT areas of the companies.} Therefore, we further separate users in the Information and Communications Technology Professional (ICT Professional) group, which is the minor group in the Professional groups since the characteristics and tasks performed by \textit{ICT professional} participants are different from other Professional, and almost half of the users from the Professional group are work in this field.

\subsubsection{Quantitative data collection: Daily task recordings}
We collected data set of 5,989 total hourly and daily tasks and associated contextual signals. 4,309 of the 5,989 tasks were associated with the worker participants that this article reports on. The average number of responses is 107.725 per participant. The highest number of responses is 219, from a participant who is a library officer, while the lowest number of responses is 29, from a participant who works as a chef. Each participant manually annotated their tasks through the ESM and DRM. The ESM provides a mechanism for conducting in-situ samplings (e.g., using periodical and on-demand surveys through a mobile device). The DRM uses a post-hoc survey to ask participants to recollect and label a sample of their tasks, activities, and contexts. We developed an in-house application (the Sensing app~\citep{liono2019building}) to facilitate the collection of the in-situ task annotations via ESM.

\textit{ESM-based Annotation}: These annotations are based on a brief survey; to avoid interfering with everyday activities and tasks, inquiries on accomplished tasks should be concise and clear. Our ESM attempts to decrease participants' cognitive biases while minimising their reliance on their ability to recall prior experiences properly~\citep{hektner2007experience}. As a result, the ESM process generally does not contain queries inquiring about the actual task start time of the mobile user. The annotations obtained through the ESM technique are referred to in this work as ``in-situ annotations''.

\begin{figure}[!h]
  \centering
  \includegraphics[width=0.93\linewidth]{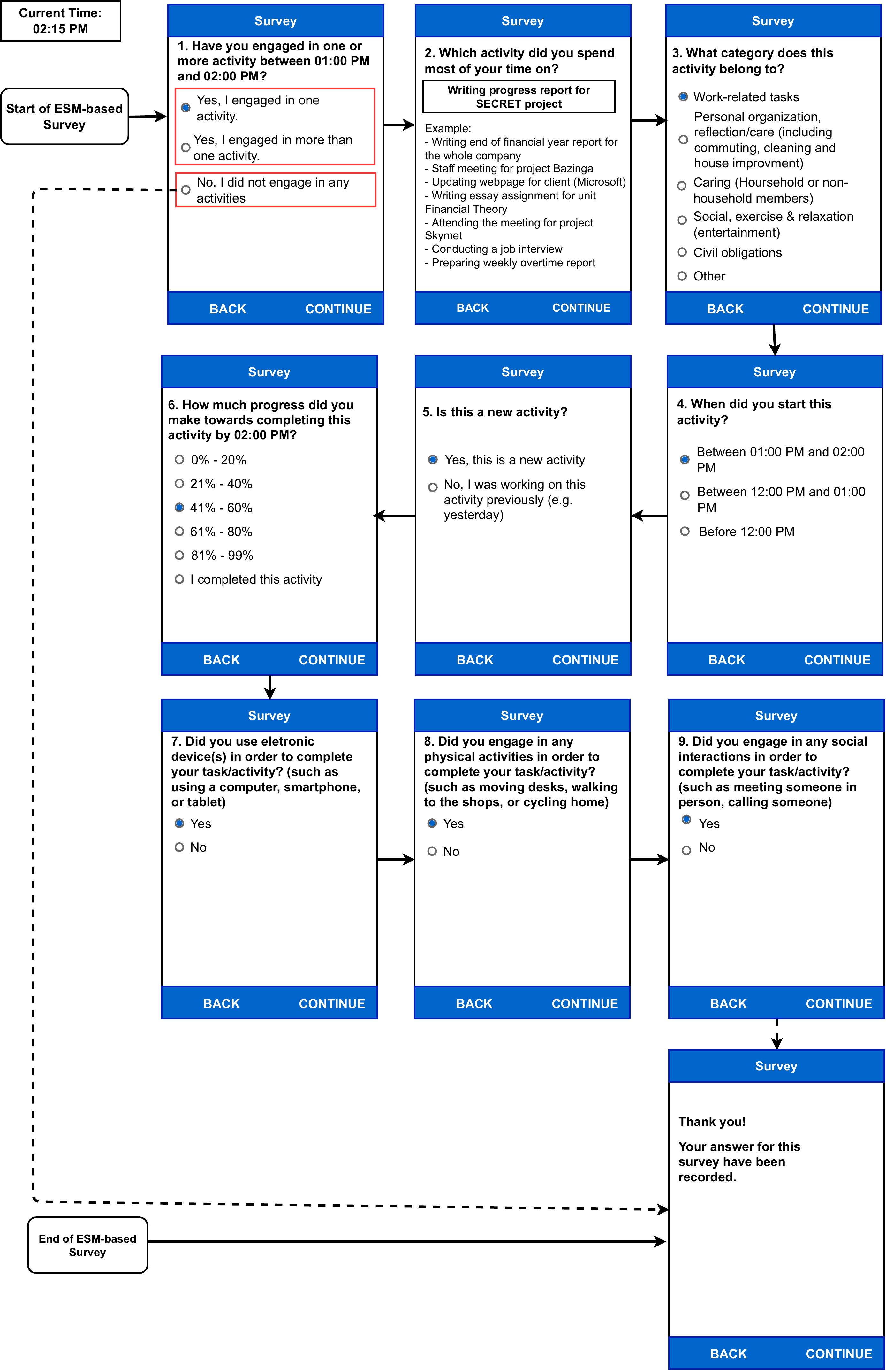}
  \caption{Workflow of ESM-based survey in the Sensing app.}
  \label{fig:ESM survey}
\end{figure}

The ESM-based survey is aimed to cover recent tasks performed by participants during the previous one-hour time block, including contextual characteristics of the activity, as seen in Figure~\ref{fig:ESM survey} for the ESM survey procedure. Assume a participant is notified to complete the survey at 2:00 PM but is unable to reply until fifteen minutes later. When the participant begins the survey at 2:15 PM, the questionnaire is restricted to tasks occurring between 1:00 PM and 2:00 PM. The contextual aspects of the reported task include 
\textit{(i)} task categorisation~\citep{trippas2019learning}, 
\textit{(ii)} the occurrence of other activities that may overlap with the reported task,
\textit{(iii)} perceived time spent, 
\textit{(iv)} a binary indication of whether this task is newly initiated or a continuation of previously reported tasks, 
\textit{(v)} estimated progress by the time block's end, and 
\textit{(vi)} the Cyber, Physical, and Social (CPS) signals presences associated with the ensuing time block. Ideally, the user should describe a recent task that may need extensive time or effort to complete. However, this ESM-based survey is provided in an open-ended format to promote flexibility and get insight into how participants view tasks in their everyday routines. The Sensing application was designed to provide hourly alerts for ESM surveys between 8:00 AM and 7:00 PM. If a participant indicates that no specific job was completed during the prior time block, the ESM-based survey will be promptly concluded. \hlnew{We use hourly alerts for ESM surveys, even though \cite{VANBERKEL2019118} indicate that an ESM survey based on smartphone screen events is more effective than a periodic survey. This is because we do not want participants to feel constantly monitored and if they miss the questionnaire corresponding to a specific time period, we can assume that they are performing an intensive task (i.e., are busy). In addition, we can conduct the cross-check with the results of the DRM survey.}

\textit{DRM-based Annotation}: Each participant was asked at the end of the day to identify all tasks completed throughout the day, including approximate start and finish times. By asking the mobile user to remember their activities within one day, we may determine the most important (and potentially time-consuming) tasks they accomplished or made progress on. The main benefit of the DRM approach is that it may be used to influence the design of future assistive technology applications by focusing on the acquisition of annotations for important tasks as viewed by users. Consider a situation in which time-consuming tasks may be assisted by an intelligent assistant's seamless recommendations of activities to enhance their completion. Inherently, annotations should contain the tasks' actual start and end times, as perceived by participants. To optimise the benefit of this approach, participants are advised to consider the DRM as a diary study, recalling the activities/tasks they had performed since waking up as a series of episodes. Retrospective task annotations are the products of this DRM-based survey procedure. The daily email alerts delivered at the end of the day (7:30 PM) serve as the trigger for this procedure in our research.

As for the contextual signals, we utilise the logging procedures to capture the contextual signals corresponding to tasks via different mobile and desktop applications, including the Sensing, RescueTime, and Journeys applications, and participants' calendar application.

\begin{itemize}
    \item[-] \textit{Sensing App}: The Sensing application logs the sensor data
from participants' Android smartphones, including internal motion unit (IMU) sensors, GPS, barometer sensor, device state, noise level, call logs, wifi scan data, Bluetooth scan data, and application usage statistics.
    \item[-] \textit{RescueTime App}: This application logs the online (cyber) behaviour of
participants. Specifically, it captures the time participants spend
interacting with different applications and websites at an hourly
granularity. These interactions are also grouped into various categories.
    \item[-] \textit{Journeys App}: This application logs sensor data from smartphone users
to detect the surrounding contexts in real-time (e.g., transport
modes, location clusters, health scores). It also provides user profiling based on historical data.
        \item[-] \textit{Calendar Events}: Events corresponding to the data collection period are extracted using the OAuth 2.0 protocol for authentication
and authorization to read their online calendar (e.g., Google Calendar). The extracted events are obfuscated with only the following
information being recorded: start/end time of an event, number
of participants, and whether a particular location was indicated.

\end{itemize}

These signals are automatically collected and paired to the task annotations based on the collected timestamp from the participant's mobile or desktop. These contextual signals contained rich information of our participants' situation when they performed different tasks. These could help us to explain the reason behind the DA's features being imagined by participants. For instance, the number of meeting events obtained from the calendar application may reflect the participant's current level of activity. As a consequence, this participant could imagine a DA which helps them manage their time by arranging meetings.

\subsubsection{Qualitative data collection: Experience-Sampling-Stimulated recall methodology (Weekly check-in)}  
In addition to recording and annotating their daily tasks, participants attended individual weekly meetings with a member of the research team. These meetings were used as a regular check-in to ensure the sensing applications used were running correctly on the individual participant's device, as well as to ensure consistency and accuracy of participant task annotations. For example, unexpected activities or gaps were identified and clarified through conversation. The initial weekly meetings (i.e., first and second weeks) allowed us to better guide our participants in understanding their tasks and improving annotations. For example, clarifying the annotation during an in-application survey as ``watching a rugby game in a stadium'' is preferred to just annotating as ``rugby'', which could then be interpreted differently as ``playing a rugby game with friends''. \hlnew{During the data collection process, we learned the interviewer should review the labeled activities prior to every meeting, in order to be more effective in our discussions. Also, any questions that may have been missed during this discussion were usually noted and were asked in the following week (since we perform weekly catch-ups, usually after their working hours).} As part of this conversation used in this work, the participant was asked to reflect on the preceding week's tasks and to describe how an imagined DA could have helped. To support the process, the participants were shown a spreadsheet containing the tasks that were logged for that week, with their annotations. This process helped the participants to recall the tasks performed and what they want DAs to support during the week.

While the conversations arising from each meeting differed based on the participant's responses, the same initial question – \textit{`imagine how a fictional or future DA could support your daily tasks based on your experience of the previous weeks'}– was used across the cohort. The use of this type of speculative questioning was informed by Luger and Sellen's \citep{luger2016like} investigation of the expectations that users had of DAs, and Meurisch and colleagues' \citep{meurisch2017reference} method of identifying needs for DA assistance. Rather than examining user expectations of an existing DA's actual abilities, however, we encouraged our participants to imagine the possible functions of a potential, or future, DA. Further, by asking participants to imagine these tasks in the context of the tasks recorded in the preceding week, our participants imagined a DA that directly responded to the requirements of their most recent work experiences. 
During these conversations, the researcher took notes that were then added to the participant's file. When taken together, these notes provide valuable qualitative insight into the tasks workers imagine a DA supporting. To understand this, and acknowledging that qualitative data requires a qualitative approach, we undertook a thematic analysis. Before describing how this analysis was undertaken, however, we offer a brief discussion of the limitations presented by this data set. 

\hlnew{Note that the process of asking participants to imagine what an `imaginary' or `future' digital assistant might do during their weekly check-ins formed part of a larger research project in which survey data and logs were used as primary data collection methods.} Given that the primary goal of this study is not to derive insights about the future DAs from these weekly notes, the nature of the data collection presented some limitations. Firstly, the data set is formed of notes arising from a conversation, rather than direct quotes arising from an in depth and structured interview. As such, these notes are in some cases the verbatim words of the participant, and in other cases they are the researcher's interpretation of the participant's responses. Given both that it is not possible to discern which is which, and the embedded nature of the researcher in both DA literature and with the participants themselves, we have treated the possible influence of this interpretation as an initial round of thematic analysis, with the participant and the researcher collaboratively identifying meaning. Secondly, although individual participants attended weekly meetings with a research team member, the notes taken during these meetings were not differentiated by time or date. It is therefore not possible to conduct an analysis of if and how our participants' imagined needs changed over time. \hlnew{Thirdly, data regarding previous use of a DA is not collected because the majority of the participants have limited experience with using DAs. Therefore, it is not possible to determine the influence of user experience on the needs of DAs.} Further, although the conversations from which these notes were made were conducted in direct relation to the hourly and daily task and sensing data described above, the notes themselves were not linked to any specific task or sensing data. As such, it is not possible to provide evidence of existing need for the imagined support described below in the task or context data. Future research that addresses each of these limitations will therefore provide a meaningful addition to the findings presented here.

\subsection{Data Analysis} 
\subsubsection{Qualitative data: Thematic analysis } 
\hlnew{Thematic analysis} is an approach to identifying descriptive themes or patterns within qualitative data \citep{braunclark}. We took an inductive approach that was driven by the data, rather than a deductive, or theory driven approach. This approach is consistent with previous research, e.g., \citep{10.1145/2987382, 10.1145/3058551, 10.1145/2491500.2491506, 10.1145/2749461}. As such, we used an open coding approach \citep{holton2007coding}. In addition to looking for patterns within the data, we also looked for their frequency. We assigned multiple codes to each group of notes associated with individual participants, and the unit of analysis was at the sentence level. \hlnew{We use the sentence as unit of analysis because the sentence is a standard unit of analysis for coding possibly multi-topic utterances in discourse analysis \citep{stojmenovic2012understanding}. As stated in Section 3.1.3, the qualitative data used for this study was obtained from researcher notes taken during weekly check-in meetings with participants. Accordingly, rather than a multi-page interview transcript, in which the unit of analysis might comprise a paragraph, or small groups of sentences, we were working with notations. To make the most of this dataset, we considered each sentence individually.} 

\hlnew{The thematic analysis} was conducted in four stages. As an example of what this multi-stage process looked like in practice, we describe below how we coded the notes resulting from all four weekly meetings associated with participant P052:
\textit{``[The future DA should] help with mundane tasks like reminders for due dates, timelines. An intelligent system that understands which reminders are heavy in expectations for completion, meaning what reminders can be ignored and which ones are too important to push to the side. An intelligent system that can find similar issues with clients from reports that are backlogged or stored in the past, from previous similar circumstances. A system that can give recommendations on what has worked and has not worked in the past.''}

\begin{table}[tb]
\small
\centering
\caption{Examples of qualitative data collected during the weekly meetings and the corresponding output of the coding process (Stage 2).}
\label{tab:code1}
\resizebox{\textwidth}{!}{\begin{tabular}{|p{10cm} | p{4.2cm} |}
\hline
\textbf{DATA}&\textbf {STAGE 2:}

\textbf {INITIAL CODE} \\
\hline

\textit{``[The future DA should] help with mundane tasks like reminders for due dates, timelines''} & Reminders and recommendation\\\hline
\textit{``An intelligent system that understands which reminders are heavy in expectations for completion, meaning what reminders can be ignored and which ones are too important to push to the side''} & Intelligent inference\\\hline
\textit{``An intelligent system that can find similar issues with clients from reports that are backlogged or stored in the past, from previous similar circumstances''} 

\textit{``A system that can give recommendations on what has worked and has not worked in the past''
} & Dynamic insights 

(past influence on future)\\\hline

\end{tabular}}
\end{table}

In the first stage, one of the researchers familiarised themselves with the data by reading and re-reading the corpus of meeting notes. 
In the second stage, the researcher coded the data, looking for trends and patterns. In this instance, the notes were coded as indicated in Table \ref{tab:code1}. In the third stage, the researcher reviewed the coded data and worked to group them into descriptive themes, as indicated in Table \ref{tab:code2}. In this example, the code `reminders and recommendation' was divided into two. `Reminders' was filed under the sub-theme scheduling, which in turn was filed under the theme `managing time'. `Recommendation' became a sub-theme that was filed under the theme `managing information'. In the last stage, we shared the themes with the broader team and reviewed them for consistency. Consistency in this context refers not to objective accuracy, but to a process of sense-checking that involved sharing and discussing the identified themes and the data underpinning them with the broader research team. In doing so, the team reviewed and confirmed that the themes were logically consistent and encapsulated the qualitative experience of the study participants. No codes were changed as a result of this stage

\begin{table}[tb!]
\small
\centering
\caption{Examples of qualitative data collected during the weekly meetings, with initial codings (Stage 2) and mappings of those codings to themes and sub-themes (Stage 3).}
\label{tab:code2}
\resizebox{\textwidth}{!}{\begin{tabular}{|p{4.9cm} | p{4.6cm}| p{4.8cm} |}
\hline
\textbf{DATA} &

\textbf {STAGE 2:}

\textbf {INITIAL CODE}&
\textbf {STAGE 3:}

\textbf {THEMES \& SUB-THEMES} \\\hline

\textit{``[The future DA should] help with mundane tasks like reminders for due dates, timelines''} &

Scheduling:

Reminders and recommendation &Theme: Managing time 

Sub-theme: Scheduling

Sub-sub-theme: Reminders
 \\\hline
 
 \textit{``An intelligent system that understands which reminders are heavy in expectations for completion, meaning what reminders can be ignored and which ones are too important to push to the side''} &
 
Intelligent inference&Theme: Managing tasks 

Sub-theme: Workflow
 \\\hline
 
 \textit{``An intelligent system that can find similar issues with clients from reports that are backlogged or stored in the past, from previous similar circumstances''} &
 
Dynamic insights 

(past influence on future )&Theme: Managing information 

Sub-theme: Recommendations
  \\\hline
 
 \textit{``A system that can give recommendations on what has worked and has not worked in the past''
}&
 
Dynamic insights 

(past influence on future)&Theme: Managing information 

Sub-theme: Recommendations
\\\hline

\end{tabular}}
\end{table}

The thematic analysis found three key themes. workers imagined a DA that could support them by managing their (1) time, (2) tasks, and (3) information. Importantly, each theme was interrelated. Managing the participants' time typically required the imagined DA to simultaneously manage the users' tasks and information. Accordingly, although the below discussion has been separated for clarity, each support area should not be viewed discretely, but rather as aspects of the interconnected responsibilities of an imagined DA.

\subsubsection{Quantitative Analysis of Self-reported Tasks from ESM, and the contextual signals }

In the second stage of analysis, we looked to see whether the quantitative sensing data collected from the participants held evidence of user need for these features. Although these participants imagined a DA that could manage their time, tasks, and information, we wanted to know whether their daily work tasks and the sensor log data indicated the existing needs for these features.

\begin{figure}[!ht]
  \centering
  \includegraphics[width=0.75\linewidth]{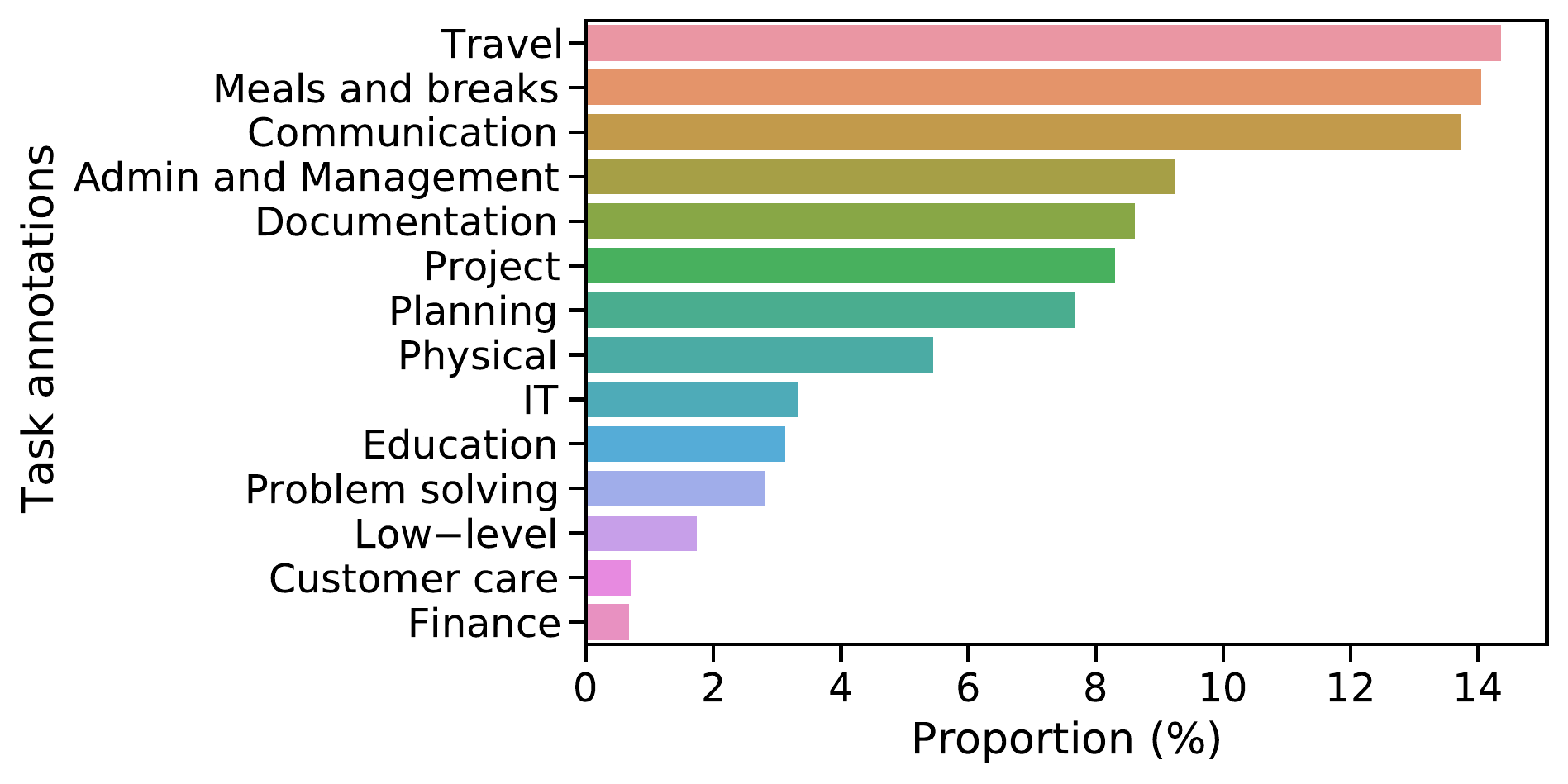}
  \caption{The proportion of task annotations in the collected dataset.}
  \label{fig:bar_task_dis}
\end{figure}

We examine the proportion of each task within the dataset. These task instances are further placed into task categories using the work-task taxonomy reported by~\citet{trippas2019learning}. 
\hlnew{To the best of our knowledge, this is the only taxonomy which focuses specifically on work-related tasks and task-performance, which is in contrast to generic time-use surveys which aims to report on how people use their time~\citep{charmes2015time}. The taxonomy's benefit is that it enables fine-grained analysis of work tasks.} The taxonomy covers the following 15 task categories: `travel', `physical', `education', `meals breaks', `communication', `planning', `project', `documentation', `low level', `admin + management', `finance', `IT', `customer care', and `problem solving'. \hlnew{We asked the participants to label their tasks with Trippas et al's taxonomy in the DRM survey (Appendix I). Then, we asked the participants to review their labels in the weekly check-in meeting before we asked them about their imagined DA.} This distribution of task categories reported by participants is illustrated in Figure~\ref{fig:bar_task_dis}. According to Figure~\ref{fig:bar_task_dis}, Travel, Meal break, and Communication are the dominant tasks reported by the 40 workers. The aggregate of these tasks accounted for approximately 40 percent of the data set. While the other task categories mostly accounted for less than 10 percent and depended on the user occupation. To investigate the relationship between features imagined by our participants and the tasks they self-reported, we apply association rule mining \citep{assoc_rule} to discover regularities between Tasks and Sub-Themes across our data. 

Moving to the sensor log data, we investigate the sensor data that has been used for passively inferring the user's task in the prior work\citep{liono2020intelligent}. We select signals which should hold the evidence about the participants' needs from those sensors. We ended up with 13 features belonging to the following groups

\textit{Device usage}: Five features representing the number of applications launched per day of five categories: Social – as frequently checking social media applications has been previously associated with user distraction \cite{gill2012distraction,mark2017blocking}, Messaging – as an indication of personal and direct communication, Email – as an indicator of professional communication, Travel – as an indication of journey planning, and Calendar – as an indication of daily scheduling. Two other features capturing the number of times when a user unlocked the screen, as well as the duration of phone activity, during the collection time. These features indicate the general amount and frequency of phone use

\textit{Journey}: There are two features that represent the number of regular and irregular places that participants visited during data collection. Two other features are used to describe the number of long-distance and short-distance journeys participants make during the day. These features provide information regarding how often participants travel and how they travel from one place to another throughout the day.

\textit{Social}: To figure out when participants need to interact with other individuals, we examine the average number of meetings and meeting attendees per day. These characteristics provide insight into the amount of time participants need to spend working with others during the day.

\hlnewtwo{To represent each participant, the daily average of these features are computed. We chose daily features over hourly or weekly as we believed that hourly features would be too fine-grained and weekly would be too coarse-grained information. Our participants were then divided into groups based on these characteristics. To identify the clusters of participants, we apply the standard clustering algorithm, K-means. Our study uses the K-means method with standard parameters (max\_iter=300, n\_int=10, and tol=1e-4) from the Sklearn library \citep{pedregosa2011scikit}. By iteratively applying K-means to a different number of clusters, an elbow graph is created using the sums of squares within each cluster to determine the optimal number of clusters. We then manually examine the demand for each group based on its characteristics.} Our findings are summarized in the following section (Section~\ref{subsubsec:needresult}). Note that participant names and workplaces have been anonymised to preserve privacy as part of the ethics approval explained in Section~\ref{subsubsec:participant recruitment}.

\section{Results}

\subsection{Contextualised by the recently performed tasks, what type of features do workers imagine a DA for workplace should take on?}

According to the data analysis described in the previous section, we illustrate the themes of imagined features in the Table \ref{tab:themes-sub-themes}. Our findings show that there are three themes mentioned by the workers, including, managing time, managing task, and managing information. The detail of these features is discussed in the following subsection.

\begin{table}[tb!]
\small
\caption{Themes and sub-themes. Frequency denotes the number of times a theme was coded across all of the qualitative data.}
\label{tab:themes-sub-themes}
\resizebox{\textwidth}{!}{\begin{tabular}{|l|p{8.8cm}|c|}
\hline
\textbf{Theme/Sub-Theme} & \textbf{Description / User quote}  & \multicolumn{1}{c|}{\textbf{Frequency}}                                             \\ \hline

\textbf{Managing time}               & The DA should assist the user in managing their time more efficiently. & \textbf{34}                                                             \\ \hline
     Scheduling             &The DA should provide \textit{``intelligent schedule management''} (P017) through \textit{``offer[ing] anticipated tasks and eventually offer realistic time estimates for tasks for better week planning''} (P023).                               & 29                                                                      \\ \hline
     
Avoiding distractions  & The DA should support the user in avoiding distractions by \textit{``gently guid[ing them] away''} (P026). 
	P028 mention that If a DA could understand repetitive tasks, and \textit{``alert user of deviation ''}
              & 5                                                                       \\ \hline

\textbf{Managing tasks}            & The DA should support the users' task management       & \textbf{27}     \\ \hline

 Workflow               & 
 P048 mentions that \textit{``Work is linear, process based''} and the DA should \textit{``Alerts, checks and reminders for next stage in the processes''}. The DA should support the user's workflow by identifying and reminding them of tasks to be completed, grouping like tasks, and breaking down existing tasks (P023).             & 17                                                                     \\ \hline
Project management     & The DA should integrate with existing platforms and provide an overview of project progress across each (P047). Assisting user to do the next step to progress in the task: \textit{``Suggest the next thing to do''} (P024)     & 10                                                                      \\ \hline
\textbf{Managing   information}      & The DA should manage, and draw inferences from, information  & \textbf{45}                                                             \\ \hline
 User improvement       & The DA should take on a \textit{``mentoring role''} (P025) to improve the user's work habits. The DA \textit{``could learn my daily pattern and create suggestions on how to better improve my effectiveness through suggestions and reminders''} (P023).                                                                                                                               & 17                                                                      \\ \hline
Recommendations        & The DA should recommend context and location specific information (P034). P051 also mentions that the DA should \textit{``recommend actions (proactively) after reminder (e.g. calendar) has been triggered''}.   & 14                                                                      \\ \hline
Traffic \& transportation              & The DA should provide information directly related to the user's commute (P023). P017 describes that the DA \textit{``would be more helpful if the intelligent assistant can arrange the activities (commuting and planning) as the working schedules can be very hectic''}      & 14                                                                      \\ \hline

\end{tabular}}
\end{table}

\subsubsection{Managing time}  
Participants imagined a DA that would assist them in managing their own and others' time by (1) providing scheduling assistance, and (2) enabling them to avoid distractions. 

\paragraph{(1) Providing scheduling assistance}
Participants imagined a DA that supported the user in managing their time by providing what participant P017  described as \textit{``intelligent schedule management''}, and participant P005 described as \textit{``intelligent reminders''}. When asked to elaborate, these `intelligent reminders' were described as alerts that drew inferences from existing data, without the user having to put \textit{``any reminder manually [in] to [their] calendar''} (P051). According to participant P023, by providing reminders for upcoming \textit{``anticipated tasks''} and eventually \textit{``offer[ing] realistic time estimates for [upcoming] tasks''}, this type of intelligent scheduling assistance would enable \textit{``better week planning''}. For example, the imagined DA might draw on both the users' previous actions and contextual information to \textit{``offer realistic estimated time for [upcoming] tasks''} (participant P023), or to learn the users' \textit{``preferences for breaks''}. As participant P039 explained, they imagined a DA that: 
\textit{``could pick up on preferences on breaks, meaning don't schedule bookings when I'm usually having lunch, and [know that] back-to-back meeting are a burden''}.

Participant P023 suggested the data required to achieve this \textit{``intelligent schedule management''} (P017) could be gathered through notifications that asked users to record \textit{``how long they believe[d] a task will take … or record[ing] which tasks a user is working on''} and for how long. 

For some participants, this intelligent scheduling assistance was imagined as extending beyond the individual user to encompass their interactions with others such as colleagues and clients. Participant P046, for example, explained that \textit{``the problem with work is that many of the meetings are ad hoc and informal''} (P046). This informality reduced the participants' ability to plan for and manage their colleagues' expectations. Accordingly, this participant imagined a DA that could mediate these interpersonal dynamics by knowing where the user was \textit{``at with [their] work''}, and, on the basis of this information \textit{``predict when these informal meetings [with the users' colleagues] will take place''} (P036). Likewise, participant P052 described a DA that could \textit{``create a void of client meetings around upcoming important activities or tasks, such as [the] financial year change''}. This would allow participant P052 to give those important activities and tasks the attention they required. By knowing not to schedule meetings too closely together, or when the user typically had a meal, the imagined DA would intelligently help the user better manage their time.

\paragraph{(2) Enabling the user to avoid distractions}

Our participants imagined a DA that managed their time by supporting them to avoid distractions. While the distractions that required avoiding varied for each participant, these were typically described as \textit{``general interrupt[ions] to tasks such as phone buzzing, slack messages, [and] notifications''} (participant P050). 

Participant P026 imagined a DA that would help them to avoid such distractions by \textit{``gently guid[ing] [them] away from [these] distractions''}. For participant P050, the imagined DA would identify when the user was \textit{``deeply focused on something [and] then shield [them] ... from distractions''}. During these times of intense focus, the imagined DA would group and hold received notifications until the user took a scheduled break. Then, participant P050, suggested, the DA could let \textit{``all messages and notifications through''}.

\subsubsection{Managing tasks} 
Participants described an imagined DA that managed the users' tasks by supporting their (1) workflow and providing, and (2) project management services. 

\paragraph{(1) Workflow}

Participants imagined a DA that could manage their tasks by supporting their workflow. While the mechanics of how this support should be provided differed based on the participants' occupation group, this assistance point was typically described as requiring a digital assistance that could identify and remind the user of remaining tasks, identify and schedule future tasks, and group like tasks together to maximise efficiency. 

A DA that could automatically \textit{``delve deeper into the connections between tasks ... and understand how [the completion of individual tasks] could help or support other [tasks]''} (P018) was considered useful by a number of participants. As participant P018 explained, this type of support would be beneficial because their work required them to \textit{``multitask and […] shift priorities when performing tasks, which constantly occurred over a course of a day.''}. A DA that could ``batch and group'' the user's work tasks would benefit participant P026 because it would enable the user to \textit{``silo or separate the work of various jobs and studies from each other so they can be focused on one at a time with maximum efficiency.''}. 

For participant P050, this grouping of tasks would support the user to better prioritise their work and home responsibilities. For example, the imagined DA could \textit{``auto-allocate''} work tasks based on their home priorities.

In addition, our participants described an imagined DA that supported the users' workflow by \textit{``identify[ing] patterns and repetitions''} within the user's workflow to reduce the \textit{``time spent on repetitive actions''} (participant P019). Further, the imagined DA should support the users' workflow by taking their existing tasks and breaking them into smaller, more easily manageable, tasks (participant P018). Similarly, participants described an imagined DA that could support the users' workflow by automating mundane tasks. Mundane tasks that participants described as requiring automation included, for example, using information from incoming communications (for example, email) to create reminders for upcoming tasks (P023), and identify and saving important details such as new phone numbers (P021). 

As noted above, each of the three areas that workers imagined a DA could support (time, tasks, and information) were interrelated. This is particularly evident in the context of managing the user's workflow. As participant P025 explained, the imagined DA should use its understanding of the user's tasks and workflow to strategically plan their time for maximum productivity. In turn, the imagined DA should use its ability to manage the user's time to inform when and how the user should complete tasks. That is, as participant P019 and P052 explained, some tasks are time sensitive, and others are not. As such, the imagined DA should manage the user's time \textit{and} tasks by differentiating between those tasks requiring immediate attention, and those that could be completed later. Participant P004 related an example from own work experience in which such intelligent time and task management would have been useful: 
 \textit{ ``There was an occasion ... where a simple email reply (approve/reject) that was forgotten to be sent caus[ed a] loss of [more than] \$10, 000. I was overwhelmed by the nature of my work ... I forgot to reply to the email because I was handling other urgent tasks.''}

\paragraph{(2) Project management}

In addition to managing the user's tasks by supporting their workflow, our participants described an imagined DA that could automate project management processes by integrating and interacting with existing platforms and multiple users. 

For participant P020, the utility of the imagined DA was directly associated with its ability to \textit{``integrate with [existing] tools and calendar[s]''} such as messaging services such as Slack, email providers such as Outlook and Gmail, calendars, Office 365, Facebook events, and Skype. The DA's imagined integration with these platforms was described by participants as a means of enhancing their existing use by providing an overview of current tasks or projects across multiple platforms. For example, participant P038 described a DA that was able to track each project they were involved with across all platforms used to provide \textit{``a summary of what has happened that is related to the project so far''}, as well as \textit{``offer[ing] actions to progress''} the project further. In this example, the relationship between the imagined DA's task and time management described above can be clearly seen. 

Ideally, our participants suggested, the imagined DA's project management support would incorporate multiple users. As participant P038 explained, if they are working on a project with multiple participants, the DA \textit{``should allow for communication relating to that project between users''}. This imagined capacity was extended by participant P021, who described a DA that could support the user's project management by \textit{``delegat[ing] tasks to people''} and providing a project overview that identified the stage at which each project component was at (whether assigned to another user, started, or completed).

\subsubsection{Managing information}
Finally, our participants imagined a DA that could support the user by managing, and drawing inferences from existing, information by (1) enabling user improvement, (2) providing recommendations, and (3) informing traffic and transportation choices.

\paragraph{(1) User improvement}

Participants imagined a DA that would use information about the user's work habits to support their improvement. According to participant P025, the imagined DA could support the user's improvement by becoming \textit{``hyper-personalised''} to the individual's \textit{``way of working''} (P025), thereby using this familiarity to identify areas for improvement. Participant P030, for example, expressed that they tended to \textit{``focus on the more attainable task''} at the expense of other, more difficult, tasks. As such, a DA that could \textit{``predict [the users'] behaviour and alert them of task[s] that need attention, in order to not let the user fall behind''} (P030) would be of significant value. Similarly, participant P028 described an imagined DA that could \textit{``alert the user of deviation''} from the task on which they were supposed to be working. In contrast, participant P030 imagined a DA that provided rewards or encouragement when the user was particularly productive. According to participant P042, such \textit{``positive reinforcement''} would encourage the user's improvement over time. 

\paragraph{(2) Recommendations}
As part of the management of information, our participants described an imagined DA that could provide location- and context-specific recommendations. The imagined recommendations were often directly linked to the DA's management of time and tasks (described above). Participant P030 for example, imagined a DA that could observe their work tasks and provide context-specific recommendations, such as \textit{``resources useful to said project''}. Likewise, participants P031 suggested a DA that, upon their arrival in a new location, could \textit{``link [the user] to ... material, exhibitions, meeting points, etc''} (P031) that were relevant to their existing interests. 

\paragraph{(3) Traffic and transportation}
Finally, our participants imagined a DA that managed their information by providing traffic and transportation advice. For participants P016, P017, P020, P029, and P018, the imagined DA would have the capacity to provide intelligent transport planning. For example, as participant P018 explained, the DA could provide them with \textit{``intelligent notification of public transport delay[s] (and suggest transport alternatives)''.} These suggestions, according to P041, \textit{``should be delivered 30-60 minutes before the user would expect to catch the train [for example], so they can know to arrange alternative transportation''.}

As noted throughout this section, the assistance points imagined by our participants were interrelated. This is particularly the case in the context of traffic and transportation advice. For example, participant P019 imagined a DA that could combine traffic and transportation information with its understanding of the user's time. That is, the imagined DA would \textit{`` be aware of the users' location, and suggest ... when to start leaving for the next meeting''} and how best to get there (P019). In addition, the imagined DA should be able to use traffic and transportation information to manage the user's tasks. As participant P036 explained, 
\textit{``the assistant should be able to track my journey to work and if it sees that my is going to be late to a meeting. Have the assistant intelligently reschedule all of his meetings''}.  Likewise, participant P053 imagined a DA that could provide notifications to team members if a task was delayed due to the delegated user being \textit{``late due to commuting issues''}.

\subsubsection{Need characterization based on sensor signals}\label{subsubsec:needresult}
\hlcolor

We analyse the characteristics of each cluster in this section and determine the potential demand that participants in each group may have based on their characteristics. This desire is then compared with the needs of participants as derived from thematic analysis, which was discussed in the previous section.

\begin{table}[tb!]
\small
\centering
\caption{The five clusters of participants based on their contextual signals. Underlined values are those that are considerably different from the mean of all participants. These values will be used to determine the characteristics of each cluster.}
\label{tab:clustering}
\resizebox{\textwidth}{!}{
\begin{tabular}{lcccccc}
\hline

\textbf {Feature/Cluster (\# participants)} & \textbf{C1 (5)}& \textbf{C2 (10) } & \textbf{C3 (5)} & \textbf{C4 (12)}& \textbf{C5 (9)} &\textbf{All}
\\

\hline\hline

$f_{1}$ :App Launched: \# Social& 16.11& 21.21& \underline{30.47}& 12.51& \underline{5.67}& 15.76 (11.14) \\
$f_{2}$ :App Launched: \# Email& \underline{13.99}& 11.51& \underline{13.29}& 6.39& \underline{2.45}& 8.54 (5.51) \\
$f_{3}$ :App Launched: \# Messaging& \underline{34.59}& 26.19& \underline{32.50}& 20.03& \underline{5.61}& 21.67 (12.80) \\
$f_{4}$ :App Launched: \# Travel& \underline{4.24}& \underline{4.41}& 2.28& 0.45& \underline{0.08}& 2.02 (3.04) \\
$f_{5}$ :App Launched: \# Calendar& 5.37& 5.36& 5.22& 4.42& \underline{1.19}& 4.15 (3.47) \\
$f_{6}$ :Sessions: \# Sessions& 32.38& 30.68& \underline{44.91}& 29.05& \underline{11.15}& 27.86 (14.14) \\
$f_{7}$ :Sessions: Duration (min)& 193.75& 244.87& 196.65& 202.62& 251.34& 221.81 (120.39) \\
\hline
$f_{8}$ :Journey: \# Long-range journeys& 3.58& \underline{5.05}& \underline{2.55}& 2.95& 2.77& 3.45 (1.42) \\
$f_{9}$ :Journey: \# Short-range journeys& \underline{6.27}& 4.29& 3.81& 2.78& \underline{2.07}& 3.54 (1.74) \\
$f_{10}$ :Journey: visit \# Regular places& \underline{3.27}& 1.78& 0.71& 1.14& \underline{0.46}& 1.35 (1.07) \\
$f_{11}$ :Journey: visit \# Non-Regular places& 15.64& \underline{20.48}& 15.43& 10.10& 15.46& 15.13 (6.30) \\
\hline
$f_{12}$ :Social: \# meeting per day& \underline{4.26}& 0.75& \underline{5.22}& 2.97& 1.08& 2.44 (2.41) \\
$f_{13}$ :Social: \# meeting attendees& 1.50& 0.88& 1.13& \underline{5.23}& 1.25& 2.34 (3.98) \\
\hline
\hline

\end{tabular}}
\end{table}

Participants are divided into five groups. Table.\ref{tab:clustering} provides the five participant clusters and the mean values of contextual features within each cluster. Then, we analyse the features that differ from other groups and characterize each cluster in terms of these features. We discovered that clusters C1 and C3 have many meetings each day, requiring the members of these groups to plan their time efficiently in order to avoid meetings clashing. It is possible that these members might want assistance in time management or scheduling. In addition, these two clusters experience a higher rate of usage of email and messaging applications, which implies that their members communicate with others frequently. A demand derived from the characteristic should relate to the communication, such as a reminder to respond to an influential message or email. These demands correspond to the results of the thematic analysis. Four out of five members from the C1 and C3 clusters highlighted the feature of ``managing time'' (e.g., providing scheduling assistance) when asked what they would expect from their future DA.

With respect to journeys, participants in cluster C1 take a high number of short-range journeys and visit regular places, while cluster C2 takes a high number of long-range journeys and travels to non-regular locations. Members of cluster C2 who often travel to destinations that are unfamiliar to them and far from their current location should have a demand related to their journey, whether it be traffic information or information about the destination. Cluster C1 members may not experience the same level of demand as cluster C2 members since the destination location is close to them, and they tend to visit areas with which they are familiar. It seems that this sentiment is also aligned with the finding from the thematic analysis that seven out of ten participants in cluster C2 and only one out of five in cluster C1 highly value the assistance of the DA regarding traffic and transportation patterns, as well as providing information about the places they visited.

A further observation is that members of cluster C4 have a high number of attendees in their meetings. Members of this cluster are likely to collaborate with others, so their needs may be influenced by this aspect. Based on the thematic analysis, the need for a ``managing tasks'' feature that requires DAs to interact with multiple users was closest to this cluster's need. There were, however, only five of twelve members who mentioned this requirement. Perhaps this is the result of the characteristics of their work role, which do not allow them to use DAs for interpersonal interaction. 

In cluster C5, the members of the cluster have a low number of application launches in the majority of application categories. Furthermore, they appear to travel less often.

There may be a need for them to be provided with DAs to support them with their journey since they do not have much travel experience. Additionally, they may require DAs who can teach them how to use new applications or suggest which application(s) would be appropriate in a particular situation. Interestingly, this corresponds to the results of the thematic analysis because participants mention the ``Managing information'' feature, which is close to the extracted needs. Five out of nine respondents in cluster C5 mention this feature. This seems reasonable since people less frequently need to travel to a new location or perform a new task in everyday life.

In summary, we may be able to draw some conclusions from participants' contextual signals with respect to their demands. Yet, some features were unable to be accurately captured by the sensor signals, such as the ``user improvement'' feature or ``Enabling the user to avoid distractions''. In this sense, the weekly-checking interview is crucial as it allows us to determine and understand the hidden demand that arose during a short period of time while collecting data.

\color{black}

\subsection{Do these imagined features of DA differ across occupation roles?} 

\begin{figure}[h!]
  \centering
  \includegraphics[width=0.95\linewidth]{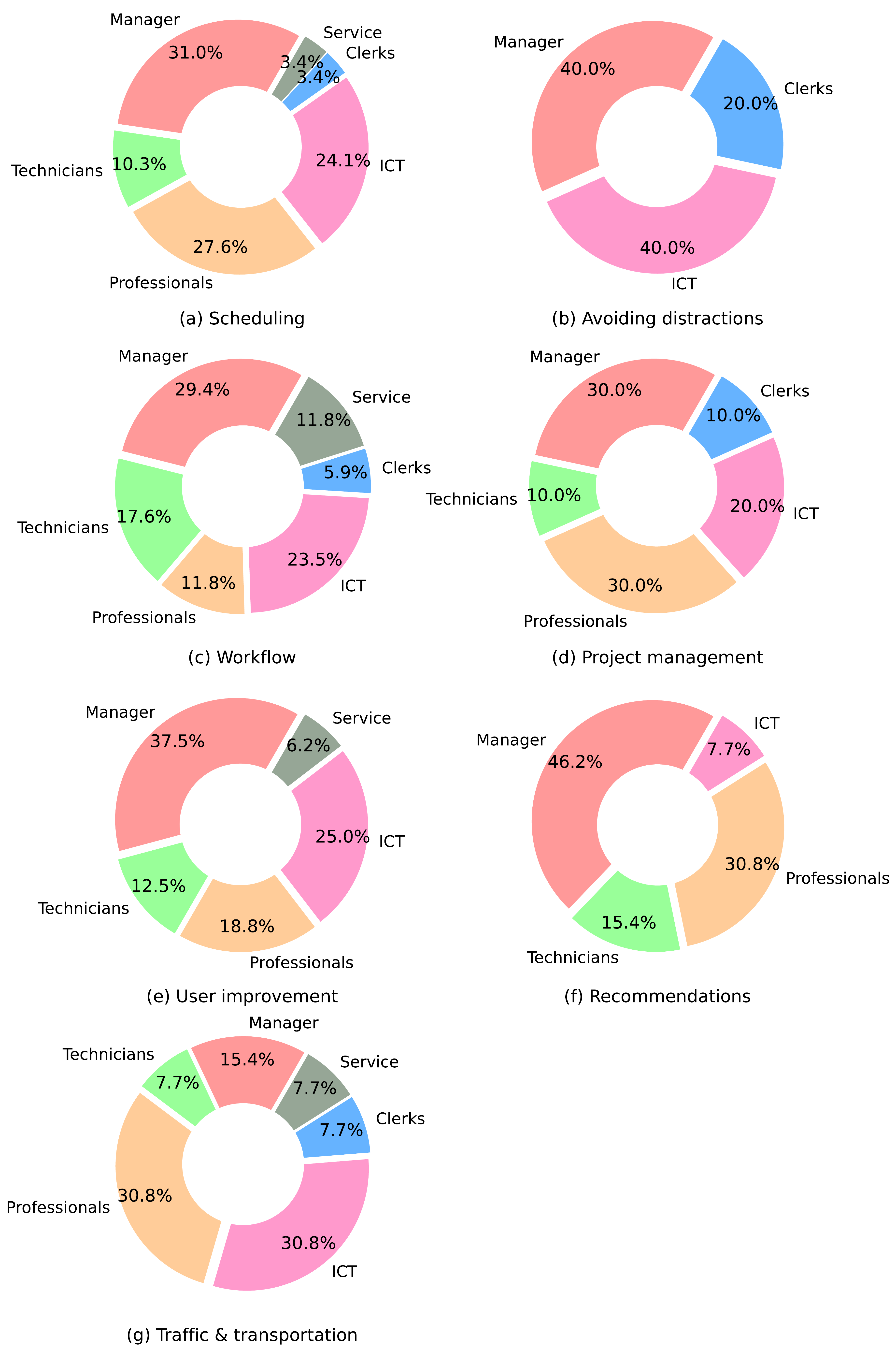}
  \caption{Proportion of occupation groups in the different sub-themes. }
  \label{fig:freq-sub-themes_t}
\end{figure}

\hlnew{Of the 40 workers involved in this research, 12 worked in \textit{Management roles}, 11 worked in the \textit{Professionals} group, 9 worked in \textit{ICT professional group}, other 4 worked in \textit{Technicians and associate professionals}, 2 worked in in \textit{Clerks group}, while the remaining 2 were employed in \textit{Service and sales workers} roles.} In this section, we address {RQ2:} Do these imagined features differ across occupation roles?

The distribution of these themes across the occupation groups is illustrated in Figure~\ref{fig:freq-sub-themes_t}. According to these figures, the imagined need for each of the identified themes differs by occupation group. For example, where Manager and ICT participants imagine a DA that can manage their time, tasks, and information, Service and sales participants do not mention project management, recommendation, and avoiding distractions at all. Moreover, avoiding distraction sub-theme was only imagined by three occupation groups. It was surprising to see that some occupations did not highlight the need for `avoiding distractions' in their wish list. This may be due to the fact that our cohort of participants only provide their responses by contextualising the tasks they had to carry out during the data collection period. Therefore, the findings indicate that the imagined features of a workplace DA can vary due to occupation groups and the tasks they need to perform as part of their work role.

\subsection{Do these imagined features relate to the tasks performed by participants?}

\begin{table}[tb!]
\small
\centering
\caption{Top 10 relationships based on the lift value from association rules between tasks and themes (all with p-value $<$ 0.1 using Fisher's exact test).}
\label{tab:assoc_rule}
\begin{tabular}{|c|c|c|}
\hline
\textbf {Antecedent (Task)} --$>$ \textbf{Consequent (Sub-Themes)} & \textbf{Lift}& \textbf{p-value}
\\
\hline
Project --$>$ Project management&2.2932&$<$ 0.0001\\ \hline 
Communication --$>$ User improvement&1.4653&$<$ 0.0001\\ \hline 
Planning --$>$ Recommendations&1.4018&$<$ 0.0001\\ \hline 
Documentation --$>$ User improvement&1.3240&0.0002\\ \hline 
Admin management --$>$ Recommendations&1.2867&$<$ 0.0001\\ \hline 
Admin management --$>$ Project management&1.2425&0.0019\\ \hline 
Documentation --$>$ Workflow&1.2395&0.0060\\ \hline 
IT --$>$ Workflow&1.2127&0.0641\\ \hline 
Project --$>$ Scheduling&1.1878&0.0003\\ \hline 
Planning --$>$ Traffic and transportation&1.1612&0.0685\\ \hline

\end{tabular}
\end{table}

In this section, we illustrate the relationship between features imagined by our participants and the tasks they self-reported. We used the pairs of sub-themes mentioned and tasks performed by each user as input for the association rule mining \citep{assoc_rule} to discover regularities between Tasks and Sub-Themes across our data. We show the top 10 relationships between tasks and sub-themes based on the lift value in Table \ref{tab:assoc_rule}. Lift measures the change (increase or decrease) in the probability of the presence of consequents with the knowledge that the antecedent is already present in the pair over the likelihood of the existence of consequents without knowing about the presence of antecedents. A lift value greater than one indicates a strong relationship between the antecedent and the consequent. The larger the lift ratio, the more significant the association. We also calculated the p-value using Fisher's exact test to quantify the statistical significance of these rules and report those values in the far-right column of Table \ref{tab:assoc_rule}. There is good sub-theme coverage in the top 10: six of the seven sub-themes of features imagined by participants in our cohort have a statistically significant (or near significant\footnote{The associations between ``Planning'' and ``Traffic and transportation'' and between ``IT'' and ``Workflow'' are significant at p=0.068 and p=0.064, respectively.}) relationship with their tasks. 

Focusing on Table \ref{tab:assoc_rule}, there are some noteworthy findings. The rules for the ``Project'' and ``Admin management'' tasks have high lift ratios with the ``Project management'' sub-theme. This means that there is a strong relationship between ``Project management'' themes emerging from the thematic analysis and tasks of that nature self-reported by the participants in our study. Workers who need to plan and administer projects would clearly benefit from project management assistance from DAs. Indeed, one of our participants remarked ``\emph{The DA should integrate with existing platforms and provide an overview of project progress across each.}'' (P047). The ``Project'' task also has a relationship with the ``Scheduling'' sub-theme; people may be more efficient and effective in performing their ``Project'' tasks if DAs can assist them with scheduling their time. The association between the ``Planning'' task and the ``Traffic and transportation'' sub-theme is also important since travel and commute information (e.g., timing, mode) may need to be considered when people plan work.

Additionally, the ``Documentation'' and ``Communication'' tasks are related to the ``User improvement'' sub-theme. One of the ``User improvement'' imagined features is a ``\emph{DA that could alert the user of deviation from the task}'' (P028); this feature is essential for ``Documentation'' tasks since workers may need to maintain attentional focus while completing tasks of this type. As for the ``Communication'' task, workers who need to communicate with many people may welcome feedback from DAs about how to do so more effectively as a part of feedback on self-improvement in general, as mentioned by participant P025: ``\emph{DA could support the user's improvement by becoming hyper-personalized to the individual's way of working and using this familiarity to identify areas for improvement}''. Moreover, the ``Planning'' and ``Admin management'' tasks are associated with the ``Recommendations'' sub-theme. This may be because some tasks (e.g., planning an exhibition event in a new place or designing a logo) require the provision of relevant material to support decision making or creativity support; if DAs can provide this information or suggest ideas, users are likely to be satisfied. Finally, there is an association between the ``Documentation'' and ``IT'' tasks and the ``Workflow'' sub-theme. These tasks may be routine or repetitive and workers may require assistance with that (e.g., one participant requested assistance with ``\emph{reducing the time spent on repetitive actions}'' (P019)) or may be complex and need to be decomposed into sub-tasks (e.g., ``\emph{DAs should understand how [to support the completion of sub-tasks]''}, imagined by P018).

Overall, this deeper understanding of the associations between the types of tasks that workers perform (from self-report data) and the specific sub-themes that were identified in their imagined features (reflecting the support required) can provide useful insights to help to inform the design of task-specific assistance in DAs. Future work will extend beyond the top 10 association rules to see if there are additional insights that can be gleaned about the task-feature relationship.

\section{Implications}

Our participants imagined a DA that supported them by managing their (1) time, (2) tasks, and (3) information. We put these findings into context, and provide some suggestions for how they might be put into practice in future DAs. 

\subsection{Time}
Our participants imagined a DA that could manage their time by providing scheduling assistance and enabling the user to avoid distractions. These findings accord with existing research. Work by \citet{cranshaw2017calendar}, \citet{berry2011ptime} and \citet{myers2007intelligent} provided valuable insight into the role that DAs can play in providing scheduling assistance. \citet{cranshaw2017calendar}, for example, describe a DA that interacts with the user's personal calendar, as well as that of their colleagues, to automatically schedule meetings. Similarly, the relationship between DAs, time management, and productivity at work has been well explored \citep{kang2017zaturi,teevan2016future,teevan2016productivity,kang2017zaturi}. \citet{kang2017zaturi}, for example, describe a mobile application that utilises the user's activity data to identify and fill what the authors define as `micro spare time'. Work by \citet{teevan2016future}, \citet{teevan2016productivity}, and \citet{white2021microtask} provided evidence for how DAs might support users to manage their time by identifying `microtasks' that enable `microproductivity'. \citet{teevan2016future} provided further evidence for our argument that the areas our participants imagined a DA should support are interconnected. \citet{teevan2016productivity} and \citet{white2021microtask} focused on how a DA might manage the user's time by identifying `microtasks' and supporting task completion. 

In addition to scheduling assistance, our participants imagined a DA that managed the user's time by supporting their avoidance of distraction. Work by~\citet{horvitz2001notification} and~\citet{czerwinski2000instant} provided valuable insight into the consequence of digital distraction, demonstrating that this concern extends beyond our participants. Our participants specifically described a DA that assisted the user to avoid digital distraction by `shielding' them, grouping and withholding notifications to be released at a more appropriate time. Early research on this topic by \citet{mcfarlane2002comparison} identified four strategies for coordinating interruptions: immediate, negotiated, mediated, and scheduled. Where immediate interruptions are provided irrespective of the status of the current task, scheduled interruptions are provided incrementally at pre-arranged points. In turn, a mediated interruption is understood as the best possible time to interrupt the user as determined by a DA. Work by \citet{iqbal2008effects} demonstrated that mediating the delivery of notifications that might cause digital distraction  (whether at breakpoints between tasks, or at points deemed statistically likely to be `interruptible') can effectively support the user's focus and productivity. As such, \citet{iqbal2008effects} found that users experience significantly lower frustration when notifications are delivered at breakpoints, rather than when initially received. Although the scheduling of notifications at breakpoints did not necessarily increase the speed at which users returned to the task at hand, \citet{iqbal2008effects} observed that users in their study \textit{preferred} receiving notifications during these breakpoints. As they suggested, this ``strongly indicates that users would accept systems that schedule notifications at breakpoints in practice'' \citep[p.~101]{iqbal2008effects}, and aligns directly with the findings discussed in this article. Further evidencing the close ties between each of the three assistance points described above, \citet{mehrotra2016my} found that although notifications can be distracting, the users' response is determined by contextual information. That is, ``the recipient's relationship with the sender of the notification, the ongoing task's type, completion level and complexity'' \citep[p.~np]{mehrotra2016my} each determine how the user responds.

In practice, the work habits and the range of tasks performed by users are largely contextual \citep{liono2020intelligent, white2019task}. Given that the individual worker undertakes a range of tasks with different levels of attention and involvement required, a DA could consider the user's regular working habits to better manage the user's time. For example, if the user tends to schedule client meetings on Mondays, the DA might schedule reporting on Thursdays. 
\hlcolor

The possible features which are implicated in this need are listed following.

\begin{itemize}
    \item By considering a user's historic behaviours and current context, future DA's may be able to suggest which tasks a user should work on next 
    \citep{zhang2021grounded} or when to best schedule them in the day or the week \citep{white2019task}.
    \item By considering a user's schedule and contextual sensor signals, future DA may be able to alert the user when they become distracted from their task, and it can inform the user where the distraction typically occurs based on historical sensor data (e.g., noise level).
\end{itemize}

\color{black}

\subsection{Tasks}
While our participants imagined a DA that managed the user's tasks by supporting two key domains (workflow and project management), the role that DAs can play in relation to supporting tasks more generally is well covered. For example, \citet{mcgregor2017more} discussed how a DA might manage tasks associated with workplace meetings, such as taking notes and identifying and assigning resulting tasks. Similarly, \citet{gil2008towards} outlined the potential for DAs to support their users by providing ``intelligent assistance in automatically interpreting, managing, automating, and in general assisting users'' \citep[p.~329]{gil2008towards} while \citet{teevan2016future} described DAs that assist the user to better manage their time through identifying `microtasks'. In addition, \citet{white2021microtask} suggested that DAs should recommend which `microtasks' can be accomplished based on the given context. This will help people make progress on their to-dos even if they only have short free time. The possible features which are implicated in this need are listed following.

Moreover, quantifying progress made on a task is a fundamental capability of intelligent task management systems \citep{white2019taskC}. Users can, but should not need to, explicitly communicate task progress to existing systems. Cues about task progress could help future DAs to perform several actions that would address the findings presented here. \hlcolor 

Following are some possible features that are associated with this need.

\begin{itemize}
    \item Offering support that is appropriate for the current task stage, providing guidance to users about how much work has been done and how much work remains (e.g., task progress bars), and/or to update estimates of task completion time in real-time. 
    \item By monitoring a user's historical data of working on a (type of) task, a DA could estimate when the task was likely to be completed, thereby supporting the user by managing their tasks \textit{and} their time. 
\end{itemize}
\color{black}

\subsection{Information}
Finally, our participants imagined a DA that would manage the user's information. Our finding that workers imagined a DA that could provide intelligent recommendations aligns with work by \citet{mcgregor2017more} which sought to understand how a DA could make meeting notes and identify action items. \citet{mcgregor2017more} found that one of the greatest barriers to this work was the lack of contextual information. Although the proposed DA could provide a list of tasks identified throughout  meetings, without the contextual information surrounding these tasks, meeting participants were unable to complete them. As such, a DA that could \textit{both} identify and manage the user's tasks \textit{as well as} provide context- and location-specific information about those tasks would be beneficial. 
\hlcolor

These are some of the possible features associated with this need. 

\begin{itemize}
    \item Future DAs could be proactive in setting up reminders for impending tasks or rescheduling appointments when clashes or unforeseen circumstances arise. 
    
     \item A future DA should be able to detect and distinguish current and future contexts in which the user operates, as well as alert the user of upcoming disruptions in those contexts.
    \item The DA could observe if the user is or will be working in a quiet office environment or the middle of their commute and provide relevant suggestions suitable and actionable to their contexts.
     
\end{itemize}
\color{black}

Our thematic analysis of qualitative data derived from a user study of 40 participants demonstrates that workers would benefit from a workplace DA that managed their time, tasks, and information.

\section{Conclusion and Future Work}

\hlnewtwo{It should be noted that our study focuses on envisioning the future utilisation of DA in everyday life. Hence, this paper should provoke further studies and exploration based on the above discussions that aim to stimulate the imagination of DA usages at work in order to help the future design of assistive technology applications.} A significant contribution of this article is 
the insight into what workers in different roles want from a future DA. Our findings confirm prior research that shows the relevance of DAs that support people's time, tasks, and information. In focusing on user desires, rather than the provisions of existing DAs, we extend this current literature; moving away from the existing focus on the use of, and user satisfaction with, DAs, to consider user needs and future possibilities. In addition, by positioning qualitative insights into these imaginings directly with quantitative self-reported task data, we highlight the relationship between occupation roles and the tasks workers imagine a workplace DA supporting.  Our findings, therefore, extend the existing literature by demonstrating the utility of supporting user time, tasks, and information, and highlighting how different occupation roles require varied approaches to each of these domains. By developing a more nuanced understanding of what workers imagine a DA for the workplace doing, we provide valuable evidence of workers' needs \citep{maedche2019ai} to inform the ongoing development of DAs.

The specific contributions of this article are as follows.
1) \textbf{Imagined themes for DAs design:} we demonstrated the perceived utility of DAs that support user time, tasks, and information by target users.
2) \textbf{Difference of needs based on occupation:} we highlighted how different occupation roles require varied approaches to each of these support domains. 3) \textbf{Effect of task characteristics:} we illustrated the effect of task characteristics on the DA features that users desire.
4) \textbf{Implications for the design of future DAs:} our findings are synthesised and implications for DA design are discussed in the context of existing literature.

Drawing on a thematic analysis of the notes arising from weekly meetings with our group of 40 worker participants, this article identifies three domains – time, tasks, and information – for consideration in the development of future workplace DAs. As we have demonstrated, each of these factors require consideration in relation to one another. A DA that manages user tasks, is also required to manage their time and information. Our findings confirm the need for designers of workplace DAs to understand multiple aspects of users' tasks, time, and information. In addition, they should also consider how different workplace roles may require different types of DA support.

The qualitative data that this article reports on forms part of a much larger project. Reassessing these qualitative findings in direct relationship to the quantitative data, i.e., the contextual signals, collected would also be of enormous benefit. \hlnew{However, there are still gaps in our findings, caused by the limitations of our data, that should be addressed in the future. First, the demands of DA users may vary according to their geographical location and time of day. Then, we will analyse how time and location influence the needs of users. Second, we demonstrate how the needs of users are related to their tasks and work roles. The experienced user, however, may conduct their work differently from the less experienced user. This may result in a difference in their expectations regarding DA support. Therefore, future studies may build upon the information provided here by documenting this influence.} Furthermore, during the COVID-19 pandemic period, people may alter their work habits. For instance, some workers may require to work from home and so stop commuting to work. The workers may be required to attend a large number of online meetings to interact with their colleagues and demonstrate their performance. This change may have an effect on the design of future DAs, since user tasks are linked to imagined features. Workers that work from home and attend many meetings may put a priority on the \textit{Time} features. As a result, the value of our identified themes for DAs employed during COVID-19 or in a post-COVID environment, would be worth further investigation. \hlnew{Finally, future research could meaningfully build on the findings presented here by taking up the use of more creative methods such as a workshop with speculative design activities or probes to address limitations of the data used in this work.}

\section*{Acknowledgment}
This research was supported by Microsoft Research through the
Microsoft-RMIT Cortana Intelligence Institute. Any opinions, findings, and conclusions expressed in this article are those of the authors and do not necessarily reflect those of the sponsor.

\newpage




 \bibliography{mybib}






\hfill

\appendix 

\textbf{Appendix I}

\textbf{A DAILY RECONSTRUCTION METHOD (DRM): }
 
\textbf{ SURVEY}

Considering a specific task \textbf{[XYZ]}, the questions in DRM-based survey would be presented as follows: 

\textbf{Q1.} What time did you wake up today? (hh:mm) 

\textbf{Q2.} How many hours did you spend for sleeping (in total)?

\begin{tabular}{l l }
• More than 8 hours & • 6 hours \\
• 8 hours  & • 5 hours \\
• 7 hours  & • Less than 5 hours \\
\end{tabular}

\textbf{Introduction to rest of survey: }
\begin{itemize}
    \item [] Thinking about today, we'd like you to reconstruct what your day was like, as if you were writing in your diary.
    \item [] Think of your day as a continuous series of scenes or episodes in a film. Each episode is a task that you have performed or in progress towards the completion.
    \item[] Each task should at least be performed in one-hour duration. In this study, we aim to understand how an intelligent assistant can help in recognizing and managing your daily tasks, to increase the overall productivity and your quality of life. \textbf{Next » }
\end{itemize}

\textbf{Q3.} Have you attempted/progressed on any tasks today? 
\begin{itemize}
    \item [•] Yes » \textbf{Proceeds to Q4.}
    \item [•] No » \textbf{Finishes the survey.}
\end{itemize}

\textbf{Q4.} Please enter the description of one task you attempted/progressed on today. 

\textbf{Q5.} To which category does \textbf{[XYZ]} belong to? 
\begin{itemize}
    \item [•] Work-related tasks 
\item [•]  Personal organization, reflection or care (includes commuting, cleaning and house improvement) 
\item [•]  Caring (household or non-household members) 
\item [•]  Social, exercise \& relaxation (entertainment) 
\item [•]  Civil obligations 
\item [•]  Other: $\_\_\_\_\_\_\_\_\_\_\_\_\_\_\_\_\_\_\_\_\_\_\_$
    
\end{itemize}

\textbf{Q6.} To which of the activity/task-type does \textbf{[XYZ]} belong to?

\begin{tabular}{l l }
• Communication & • Problem solving \\
• Documentation & • Low-level \\
• Planning & • Project \\
• Admin and management & • Customer care \\
• Education & • Meals and breaks\\ 
• IT & • Travel \\
• Finance & • Other: $\_\_\_\_\_\_\_\_\_\_\_\_\_\_\_\_\_\_\_\_\_\_\_$ \\
• Physical
\end{tabular}

\textbf{Q7.} What kind of trigger did you initiate \textbf{[XYZ]}? 

\begin{itemize}

\item [•] Deadline 
\item [•] Reminder/alarm (e.g. through digital notification) 
\item [•] Ad-hoc/spontaneously 
\item [•] Needs for resources 
\item [•]  Other: $\_\_\_\_\_\_\_\_\_\_\_\_\_\_\_\_\_\_\_\_\_\_\_$

 \end{itemize}

\textbf{Q8.} What is the approximate time when you started \textbf{[XYZ]} (hh:mm format)?

\textbf{Q9.} Approximate progress when you started Schedule a Meeting: 

\begin{tabular}{l l }
• 0\% – 19\% & • 80\% – 99\% \\
• 20\% – 39\% & • 80\% – 99\%  \\
• 40\% – 59\% & • 100\% (complete) \\
• 60\% – 79\%  

\end{tabular}

\textbf{Q10.} What is the approximate time when you stopped \textbf{[XYZ]} (hh:mm format)?

\textbf{Q11.} Approximate progress when you stopped \textbf{[XYZ]}: 

\begin{tabular}{l l }
• 0\% – 19\% & • 80\% – 99\% \\
• 20\% – 39\% & • 80\% – 99\%  \\
• 40\% – 59\% & • 100\% (complete) \\
• 60\% – 79\%  

\end{tabular}

\textbf{Q12.} How satisfied are you with the progress of \textbf{[XYZ]}? 

\begin{itemize}
\item [•] Extremely satisfied 
\item [•] Somewhat satisfied 
\item [•] Neither satisfied nor dissatisfied 
\item [•] Somewhat dissatisfied 
\item [•] Extremely dissatisfied 

\end{itemize}

\textbf{Q13.} Thinking about the urgency of this task, what was your perceived priority of when you started \textbf{[XYZ]}: 
\begin{itemize}
\item [•] High 
\item [•] Medium 
\item [•] Low  
\end{itemize}

\textbf{Q14.} Who were you directly interacting with in the progression of \textbf{[XYZ]}?

\begin{tabular}{l l }
• None & • Co-worker(s) \\
• Spouse/significant other & • Boss(es) \\
• Household member(s) & • Other: $\_\_\_\_\_\_\_\_\_\_\_\_\_\_\_\_\_\_\_\_\_\_\_$ \\
• Friend(s) 

\end{tabular}

\textbf{Q15.} Describe your activities and contexts involved for the progression of \textbf{[XYZ]}. 

\textbf{Q16.} \textbf{[XYZ]}? Recalling today's tasks, is there any more task you attempted to progress on? 

\begin{itemize}
\item [•]  Yes » \textbf{Loops back to Q4.}
\item [•]  No » \textbf{Finishes the survey.}

\end{itemize}
\end{document}